\documentstyle[aps,preprint,epsfig]{revtex}
\begin{document}
\title
{$b{\bar b}b{\bar b}$ production in proton-nucleus collisions at
the LHC}
\author{A. Del Fabbro and D.Treleani}
\address{ Dipartimento di Fisica Teorica dell'Universit\`a di Trieste and
INFN, Sezione di Trieste,\\ Strada Costiera 11, Miramare-Grignano,
I-34014 Trieste, Italy.} \maketitle
\begin{abstract}
A sizable rate of events, with several pairs of $b$-quarks
produced contemporarily by multiple parton interactions, may be
expected at very high energies as a consequence of the large
parton luminosities. The production rates are further enhanced in
hadron-nucleus reactions, which may represent a convenient tool to
study the phenomenon. We compare the different contributions to
$b{\bar b}b{\bar b}$ production, due to single and double parton
scatterings, in collisions of protons with nuclei at the CERN-LHC.
\end{abstract}
\vspace{3cm}
E-mail delfabbr@trieste.infn.it \\
E-mail daniel@trieste.infn.it \\
\newpage
\section{Introduction}

The large rates of production of heavy quarks, expected at high
energies, may lead to a sizable number of events, at the CERN-LHC,
containing two or more pairs of $b$ quarks, generated
contemporarily by independent partonic collisions. An inclusive
cross section of the order of $10$ $\mu$b may in fact be foreseen
for a double parton collision process, with two $b{\bar b}$ pairs
produced in a $pp$ interactions at $14$TeV, while the cross
section to produce two $b{\bar b}$ pairs by a single collision at
the same c.m. energy may be one order of magnitude
smaller\cite{DelFabbro:2002pw}. All production rates are
significantly enhanced in proton-nucleus collisions, which may
offer considerable advantages for studying multiparton collision
processes\cite{Strikman:2001gz}. Given the large rates expected,
the production of multiple pairs of $b$-quarks should hence
represent a convenient process to study multiparton interactions
in $pA$ collisions at the LHC. On the other hand the mechanism of
heavy quarks production is not yet understood satisfactorily also
in the simplest case of nucleon-nucleon collisions, the effects of
higher order corrections in $\alpha_s$ being still a matter of
debate. A comprehensive description, of the much more structured
process of heavy quarks production in hadron-nucleus interactions,
might hence be approached after gaining a deeper understanding of
the short scale parton-level dynamics of the process. On the other
hand a significant feature of higher order corrections in
$\alpha_s$ is that, for a limited set of physical observables, the
whole effect of higher orders reduces to an approximate rescaling
of a lowest order calculation of heavy quarks production in
perturbative QCD\cite{Ryskin:2000bz}. Some of the features of the
process are therefore effectively described by the simplest parton
level dynamics at the lowest order in $\alpha_s$, which lets one
speculate that a similar property might hold also for a much more
complex process, as $b{\bar b}b{\bar b}$ production.

Taking this optimistic point of view we will attempt, in the
present note, to obtain indications on some properties of $b{\bar
b}b{\bar b}$ production in $pA$ interactions at the LHC, by
considering the contributions of the two different interaction
mechanisms, the connected $2\to4$ and the disconnected $(2\to2)^2$
parton processes, which will be effectively described by the
lowest order diagrams in perturbative QCD, keeping higher order
corrections into account by a simple overall rescaling. Obviously,
following this philosophy, all considerations will be limited to
the class of physical observables falling in the category
mentioned above.

\section{$b\bar{b}b\bar{b}$ production in proton-nucleus collisions}

Quite in general\cite{Paver:1982yp}\cite{Braun:2000ua}, with the
only assumption of factorization of the hard component of the
interaction, the expression of the double parton scattering cross
section to produce two $b{\bar b}$ pairs is given by

\begin{equation}
\sigma_{(p,A)}^D(b{\bar b},b{\bar b})
=\frac{1}{2}\sum_{ij}\int\Gamma_p(x_i,x_j;s_{ij})
\hat{\sigma}(x_i,x_i')
\hat{\sigma}(x_j,x_j')\Gamma_{(N,A)}(x'_j,x'_j;s_{ij})dx_idx_i'dx_jdx_j'd^2s_{ij},
\label{sigmaD}
\end{equation}
where the index $N$ refers to nucleon and $A$ to a nucleus, while
the indices $i,j$ to the different kinds of partons that
annihilate to produce a $b{\bar b}$ pair and the factor $1/2$ is a
consequence of the symmetry of the expression for exchanging $i$
and $j$. The interaction region of a hard process is very small as
compared to the hadron scale, so the two interactions are well
localized in transverse space within the two overlapping hadrons.
The rate of events where two hard collisions take place
contemporarily, in a given inelastic hadron-hadron interaction,
depends therefore on the typical transverse distance between the
partons of the pairs undergoing the multiple collision processes.
A main reason of interest is hence that double parton scatterings
may provide information on the typical transverse separation
between pairs of partons in the hadron structure. Indeed the
non-perturbative input of a double parton collision is the
two-body parton distribution function $\Gamma(x_1,x_2,s_{1,2})$,
which depends not only on the fractional momenta $x_{1,2}$, but
also on the relative distance in transverse space $s_{1,2}$,
besides (although not written explicitly to simplify the notation)
on the scales of the two interactions and on the different kinds
of partons involved. As a consequence the double parton scattering
cross section is characterized by a linear dependence on
dimensional scale factors, which have a direct relation with the
typical transverse distances between the partons of the various
pairs, contributing to the double scattering process under
consideration.

The cross section is simplest when the target is a nucleon and
partons are not correlated in fractional momenta, which may be a
sensible approximation in the limit of small $x$. In such a case
the two-body parton distribution may be factorized as
$\Gamma_p(x_i,x_j;s_{ij})=G(x_i) G(x_j) F(s_{ij})$, where $G(x)$
are the usual one-body parton distributions and $F(s)$ a function
normalized to 1 and representing the parton pair density in
transverse space. The inclusive cross section to produce two pairs
of $b$ quarks is then written as\cite{Calucci:1999yz}
\begin{equation}
\sigma^D_N(b{\bar b};b{\bar
b})=\frac{1}{2}\sum_{ij}\Theta^{ij}\sigma_{i}(b{\bar
b})\sigma_{j}(b{\bar b}) \label{sdouble}
\end{equation}
where $\sigma_{i}(b{\bar b})$ represents the inclusive cross
sections to produce a $b{\bar b}$ pair in a hadronic collision,
with the index $i$ labelling a definite parton process. The
factors $\Theta^{ij}$ have dimension an inverse cross section and
result from integrating the products of the two-body parton
distributions in transverse space. In this simplified case the
factors $\Theta^{ij}$ provide a direct measure of the different
average transverse distances between different pairs of partons in
the hadron structure \cite{Calucci:1999yz}
\cite{DelFabbro:2000ds}.

The cross section has a more elaborate structure in the case of a
nuclear target. The most suitable conditions to study the
phenomenon are those where the nuclear distributions are additive
in the nucleon parton distributions. In such a case one may
express the nuclear parton pair density,
$\Gamma_{A}(x'_j,x'_j;s_{ij})$, as the sum of two well defined
contributions, where the two partons are originated by either one
or by two different parent nucleons:

\begin{equation}
\Gamma_{A}(x'_i,x'_j;s_{ij})=\Gamma_{A}(x'_i,x'_j;s_{ij})\Big|_{1}+
\Gamma_{A}(x'_i,x'_j;s_{ij})\Big|_{2}\label{gamma12}\end{equation}
and correspondingly $\sigma_D^A=\sigma_D^A|_1+\sigma_D^A|_2$. The
two terms $\Gamma_A|_{1,2}$ are related to the nuclear nucleon's
density. Introducing the transverse parton coordinates
$B\pm\frac{s_{ij}}{2}$, where $B$ is the impact parameter of the
hadron-nucleus collision, one may write

\begin{equation}
\Gamma_{A}(x'_i,x'_j;s_{ij})\Big|_{1,2}=\int
d^2B\gamma_A\Big(x'_i,x'_j;B+\frac{s_{ij}}{2},B-\frac{s_{ij}}{2}\Big)\Big|_{1,2}
\end{equation}
where $\gamma_A|_{1,2}$ are given by
\begin{eqnarray}
&&\gamma_A\Big(x'_i,x'_j;B+\frac{s_{ij}}{2},B-\frac{s_{ij}}{2}\Big)\Big|_{1}=
\Gamma_N(x'_i,x'_j;s_{ij})T(B)\cr
&&\gamma_A\Big(x'_i,x'_j;B+\frac{s_{ij}}{2},B-\frac{s_{ij}}{2}\Big)\Big|_{2}=G_N(x'_i)G_N(x'_j)
T\Big(B+\frac{s_{ij}}{2}\Big)T\Big(B-\frac{s_{ij}}{2}\Big)
\label{gammaa}
\end{eqnarray}
with $T(B)$ is the nuclear thickness function, normalized to the
atomic mass number $A$ and $G_N$ nuclear parton distributions
divided by the atomic mass number.

In the simplest additive case, the first term in
Eq.(\ref{gamma12}) obviously gives a simple rescaling of the
double parton distribution of an isolated nucleon
\begin{equation}
\Gamma_{A}(x'_i,x'_j;s_{ij})\Big|_{1}=\Gamma_N(x'_i,x'_j;s_{ij})\int
d^2BT(B) \end{equation} and the resulting contribution to the
double parton scattering cross section is the same as in a
nucleon-nucleon interaction, apart from an enhancement factor due
to the nuclear flux, which is given by the value of the atomic
mass number $A$:
\begin{equation}
\sigma_A^D\Big|_1=A\sigma_N^D. \label{sigma1}
\end{equation}
The $\sigma_A^D|_2$ term has more structure. In this case the
integration on the relative transverse distance between the
partons of the interacting pairs, $s_{ij}$, involves both the
projectile and two different target nucleons:

\begin{equation}
\int ds_{ij}\Gamma_p(x_i,x_j;s_{ij})T\Big(B+
\frac{s_{ij}}{2}\Big)T\Big(B-\frac{s_{ij}}{2}\Big). \label{gt}
\end{equation}
As one may notice in this case two very different scales are
compared, the hadron radius $r_p$ and the nuclear radius $R_A$. A
usual approximation in $pA$ interactions is to consider the limit
$r_p\ll R_A$, where one may use the approximation
\begin{equation}
T\Big(B\pm\frac{s_{ij}}{2}\Big)\simeq T(B), \end{equation} which
allows one to decouple the integrations on $s_{ij}$ and on $B$.
One hence obtains:

\begin{equation}
\sigma_A^D\Big|_2=\frac{1}{2}\sum_{ij}\int
G_p(x_i,x_j)\hat{\sigma}(x_i,x_i')
\hat{\sigma}(x_j,x_j')G_N(x_i')G_N(x_j')dx_idx_i'dx_jdx_j'\int
d^2BT^2(B), \label{sigma2}
\end{equation}
where
\begin{equation}
G_p(x_i,x_j)=\int d^2s_{ij}\Gamma_p(x_i,x_j;s_{ij}).\end{equation}

Remarkably the two terms $\sigma_A^D\Big|_1$ and
$\sigma_A^D\Big|_2$ have very different properties. In fact the
correct dimensionality of $\sigma_A^D\Big|_1$ is provided by
transverse scale factors related to the {\it nucleon} scale, cfr.
Eq.s(\ref{sdouble}),(\ref{sigma1}). The analogous dimensional
factor in $\sigma_A^D\Big|_2$ is provided by the {\it nuclear}
thickness function, which is at the second power, being two the
target nucleons involved in the interaction.

As pointed out in ref.\cite{Strikman:2001gz}, while on general
grounds $\sigma^D_{(p,A)}$ depends both on the longitudinal and
transverse parton correlations, the $\sigma_A^D\Big|_2$ term
depends solely on the longitudinal momentum fractions $x_i,x_j$ so
that, when the $\sigma_A^D\Big|_2$ term is isolated, one has the
capability of measuring the longitudinal and, a fortiori, also the
transverse parton correlations of the hadron structure in a model
independent way.

Although the two contributions may be defined also in a more
general case, the separation of the cross section in the two terms
$\sigma_A^D\Big|_1$ and $\sigma_A^D\Big|_2$ is most useful in the
regime of additivity of the nuclear structure functions, which may
not be a bad approximation for a sizable part of the kinelatical
regime of bottom quarks production at the LHC. In the case of a
central calorimeter with the acceptance of the ALICE detector
($|\eta|<0.9$), the average value of momentum fraction of the
initial state partons, in a $pp\to b{\bar b}b{\bar b}$ process, is
$\langle x\rangle\approx 6\times10^{-3}$. By introducing a cut in
the transverse momenta of the $b$ quarks of $5$ GeV one obtains
$\langle x\rangle\approx 10^{-2}$, while a cut of $20$ GeV in
$p_t$, within the same pseudorapidity range, gives $\langle
x\rangle\approx 2\times10^{-2}$. If considering a more forward
detector, as LHCb ($1.8<\eta<4.9$), the average value of momentum
fraction is $\langle x\rangle\approx 5\times10^{-2}$. Deviations
from additivity at low $x$ are less than $10\%$ for
$x\ge2\times10^{-2}$\cite{Amaudruz:1995tq} and, although
increasing with the atomic mass number, non additive corrections
are at most a $20\%$ effect, on the kinematical regime above.

As already mentioned, heavy quarks production is characterized by
a non trivial dynamics, in such a way that also next-to-leading
corrections to the lowest order term in perturbative QCD are not
sufficient for an exhaustive description of the inclusive spectra,
which most likely need an infinite resummation to be evaluated. By
comparing the results of different approaches to heavy quarks
production (as NLO QCD and $k_t$-factorization) one nevertheless
finds that, in a few cases, the whole effect of higher orders is
to a large extent just an approximate rescaling of the results
obtained by a lowest order evaluation in perturbative QCD. When
limiting the discussion to an accordingly restricted set of
physical observables, the whole effect of higher order corrections
to heavy quark production is hence summarized by a single number,
the value of the so-called $K$-factor. By evaluating, with the
$k_t$-factorization approach, rapidity and pseudorapidity
distributions of $b{\bar b}$ production, in $pp$ collisions at the
center-of-mass energy of ${\sqrt s}=5.5$ TeV, within $|\eta|<0.9$ and $1.8<\eta<4.9$, with different cuts
in the transverse momenta of the produced $b$ quarks, one finds a
result not incompatible with a lowest order calculation in
perturbative QCD rescaled by factor $K=5.5$.

To discuss the production of $b{\bar b}b{\bar b}$ in $pA$
collisions, while remaining in a kinematical regime where non
additive corrections to the nuclear structure functions are not a
major effect, we have limited all considerations to physical
observables, where higher orders may be taken into account by the
simple rescaling of the lowest order calculation. For $ b{\bar
b}b{\bar b}$ production, where only results of three level
calculations are up to now available, we have further assumed that
the value of the $K$-factor in the $2\to4$ parton process is the
same as in the $2\to2$ process. We have hence evaluated the
various contributions to the cross section in the case of a
central calorimeter with the acceptance of the ALICE detector,
$|\eta|<0.9$, and of a more forward detector as LHCb,
$1.8<\eta<4.9$. In both cases the typical values of $x$ are rather
small, so the factorization of the cross section in
Eq.(\ref{sdouble}) may not be unreasonable. Since the process is
dominated by gluon fusion, the expression of the cross section in
Eq.(\ref{sdouble}) may be limited to a single term only. For the
corresponding dimensional scale factor we have used the value
reported by the CDF measurement of double parton
collisions\cite{Abe:1997bp}\cite{Abe:1997xk}. To obtain the cross
section at the lowest order in pQCD we have used the MRS99 parton
distributions\cite{mrs99}, with factorization and renormalization
scale equal to the transverse mass of the $b$-quark. The cross
sections of the $2\to4$ processes have been evaluated by
generating the matrix elements of the partonic amplitudes with
MadGraph~\cite{madgraph} and HELAS~\cite{helas}. For the mass of
the bottom quark we have used the value $m_b=4.6$ GeV. The
multi-dimensional integrations have been performed by
VEGAS~\cite{vegas} and all lowest order pQCD cross section have
been finally multiplied by a factor $K=5.5$.

\section{Results}

A major result of the present analysis is that the effects induced
by the presence of the nucleonic degrees of freedom, in double
parton scatterings with a nuclear target, cannot be reduced to the
simple shadowing corrections of the nuclear parton structure
functions, whose effect is to decrease the cross section as a
function of $A$. In the case of double parton collisions, the main
effect of the nuclear structure is represented by the presence of
the $\sigma_A^D|_2$ term in the cross section, which scales with a
different power of $A$ as compared to the single scattering
contribution, producing an additive correction to the cross
section.

\begin{figure}
\vspace{.1cm} \centerline{ \epsfysize=8cm \epsfbox{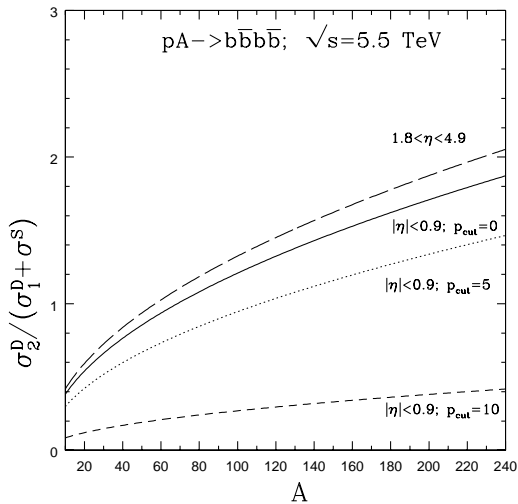}}
\vspace{.1cm} \caption[ ]{\footnotesize Relative weights of the
terms with ``anomalous'' and ``usual'' $A$-dependence in the double
scattering cross section for $b\bar b b \bar b$ production.}
\label{total}
\end{figure}

To emphasize the resulting ``anomalous'' dependence of the double
parton scattering cross section, as a function of the atomic mass
number, we have plotted in fig.\ref{total} the ratio
$\sigma_2^D/(\sigma_1^D+\sigma^S)$, as a function of $A$. The
ratio represents the contribution to the cross section of the
processes where two different nuclear target nucleons are involved
in the interaction, scaled to the contribution where only a single
target nucleon is involved. The dependence on the atomic mass
number of the latter terms of the cross section, namely the single
($\sigma^S$) and the double ($\sigma_1^D$) parton scattering terms
against a single nucleon in the nucleus, is the same of all hard
processes usually considered, where nuclear effects may be wholly
absorbed in the shadowing corrections to the nuclear structure
functions. The contribution to the cross section of the
$\sigma_2^D$ term is, on the contrary, ``anomalous'', involving two
different target nucleons in the interaction process. The ratio
above hence represents the relative weights of the ``anomalous'' to
the ``usual'' contributions to the double parton scattering cross
section on a nuclear target. The plots in fig.\ref{total} refer to
the case of a forward calorimeter ($1.8\le\eta\le4.9$) and of a
central calorimeter ($|\eta|\le.9$) with different cuts on the
transverse momenta of the produced $b$-quarks ($p_{cut}=0,5,10$
GeV/c).

\begin{figure}
\begin{center}
\epsfig{figure=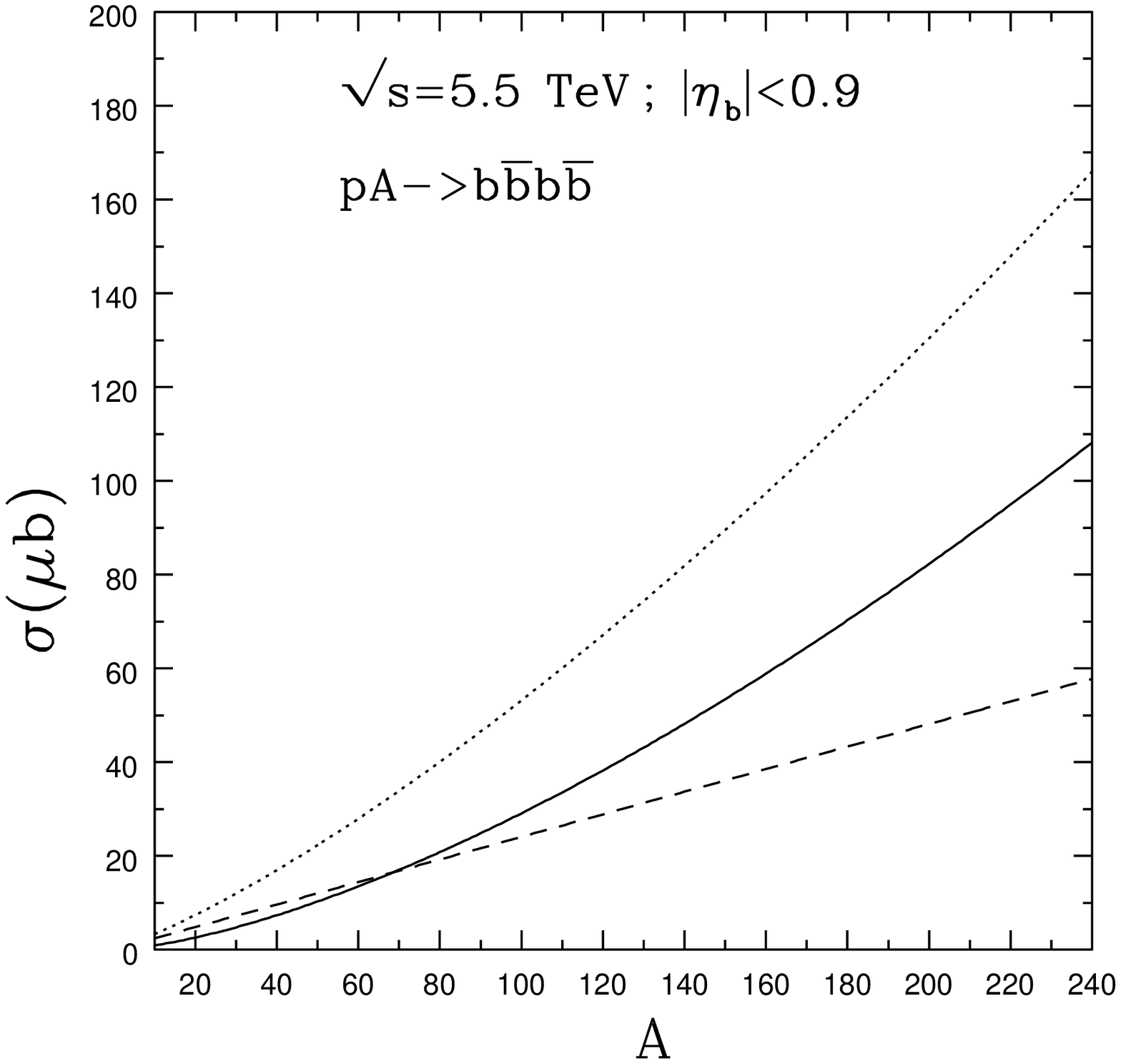,height=2.8in}
\epsfig{figure=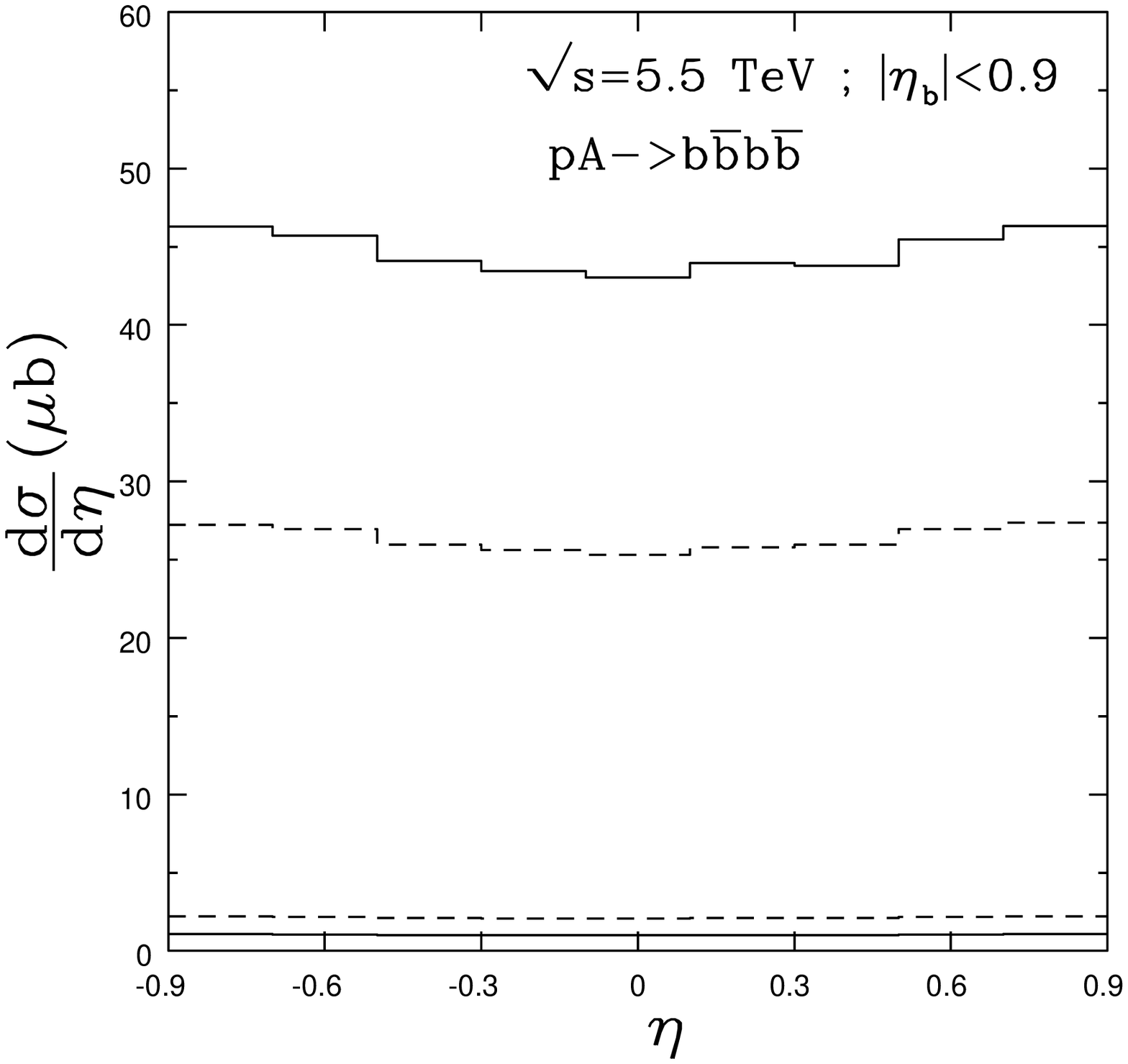,height=2.8in} \caption[ ]
{\footnotesize Different contributions to the cross section for
$b\bar b b \bar b$ production, in a central calorimeter. Left
figure: cross section as  a function of $A$ of the one-nucleon
(dashed line) and of the two nucleons (continuous line)
contributions. The dotted line is the sum of the two terms. Right
figure: differential rapidity distributions of a $b$-quark
produced in an event with two $b{\bar b}$ pairs. One-nucleon
(dashed histograms) and two-nucleon contributions (continuous
histograms) in the case of a heavy (higher histograms) and of a
light nucleus (lower histograms).} \label{A-eta-central}
\end{center}
\end{figure}

The different contributions to the cross section for $b\bar b b
\bar b$ production, due to interactions with a single or with two
different target nucleons, are shown in fig.\ref{A-eta-central} in
the case of a central calorimeter. The left figure shows the cross
section as a function of the atomic mass number of the one-nucleon
(dashed line) and of the two nucleons (continuous line)
contributions to the cross section. The dotted line is the sum of
the two terms. The right figure shows the two contributions to the
rapidity distribution of a $b$-quark produced in an event with two
$b{\bar b}$ pairs: one-nucleon (dashed histograms) and two-nucleon
contributions (continuous histograms) in the case of a heavy, {\it Au},
(higher histograms) and of a light nucleus, {\it O}, (lower histograms). 
The dependence on $\eta$ of the two histograms is essentially the
same, showing that the effect of the single parton scattering term
(the $2\to4$ parton process) is negligible in this kinematical
regime.

\begin{figure}
\begin{center}
\epsfig{figure=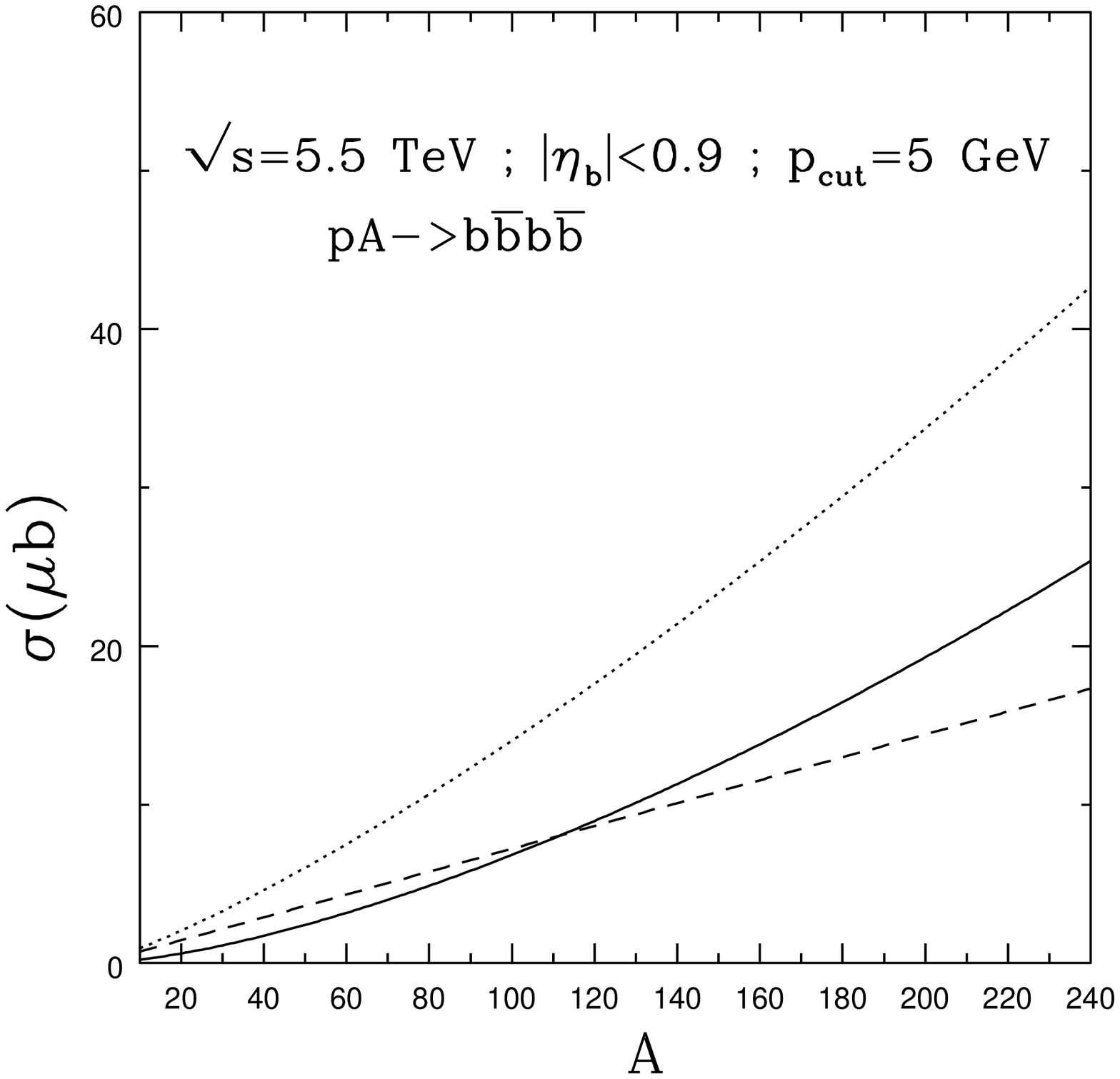,height=2.8in}
\epsfig{figure=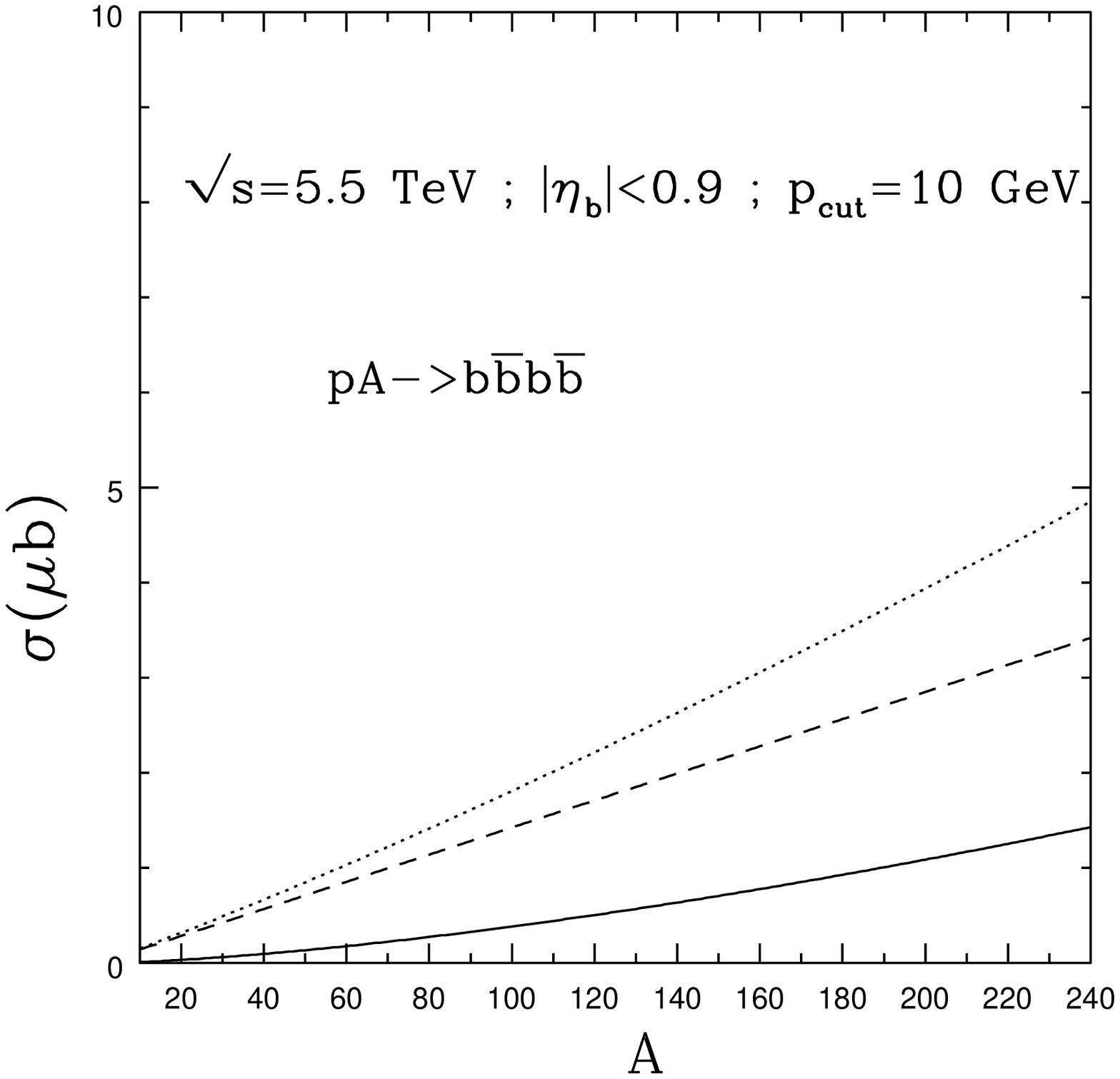,height=2.8in} \caption[ ]
{\footnotesize Different contributions to the cross section for
$b\bar b b \bar b$ production in a central calorimeter as a
function of $A$, after applying a cut of $5$ GeV (left figure) and
of $10$ GeV (right figure) in the transverse momenta of each
produced $b$-quark: one-nucleon (dashed line), two nucleons
(continuous line) processes and total (dotted line).}
\label{A-cuts}
\end{center}
\end{figure}

The $A$-dependence of the two different contributions, as a
function of $A$, are shown in fig.\ref{A-cuts} in the case of a
central calorimeter, after applying a cut of $5$ GeV/c (left
figure) and of $10$ GeV/c (right figure) in the transverse momenta
of each produced $b$-quark. Dashed, continuous and dotted lines
have the same meaning as in fig.\ref{A-eta-central}.

\begin{figure}
\begin{center}
\epsfig{figure=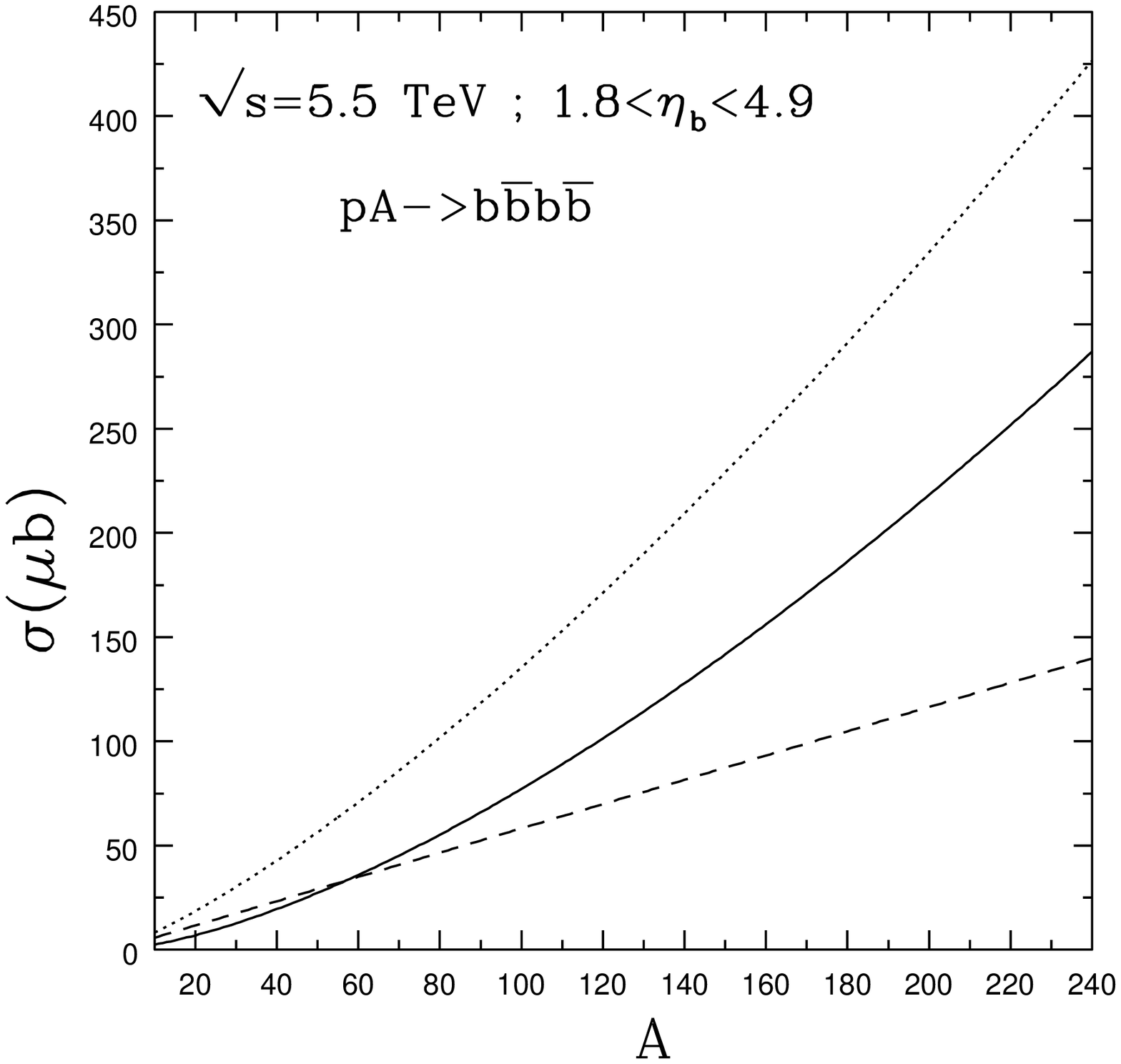,height=2.8in}
\epsfig{figure=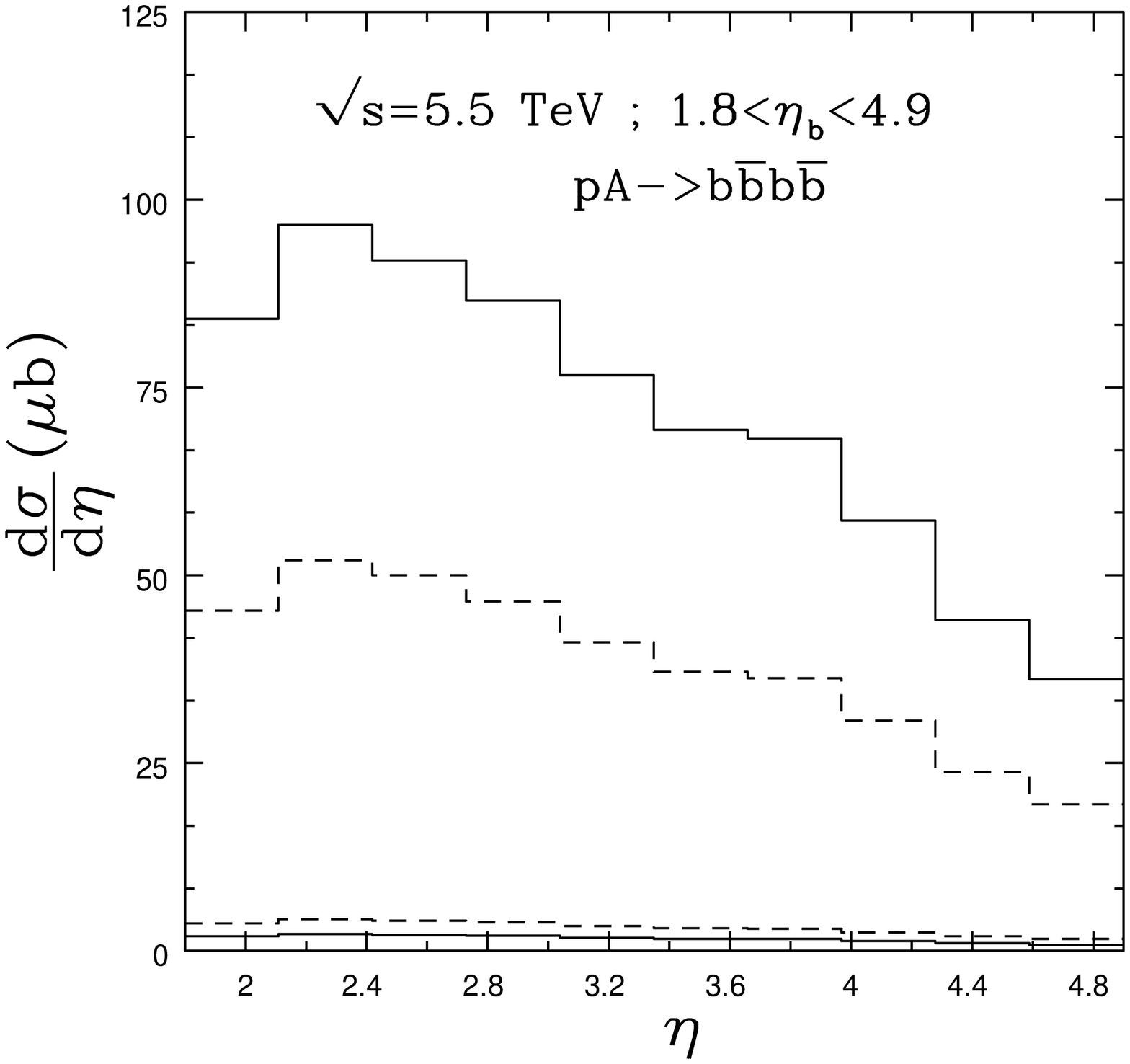,height=2.8in} \caption[ ]
{\footnotesize Different contributions to the cross section for
$b\bar b b \bar b$ production, in a forward calorimeter. Left
figure: cross section as  a function of $A$ of the one-nucleon
(dashed line) and of the two nucleons (continuous line)
contributions. The dotted line is the sum of the two terms. Right
figure: differential rapidity distributions of a $b$-quark
produced in an event with two $b{\bar b}$ pairs. One-nucleon
(dashed histograms) and two-nucleon contributions (continuous
histograms) in the case of a heavy (higher histograms) and of a
light nucleus (lower histograms).} \label{A-eta-forward}
\end{center}
\end{figure}

The case of a forward calorimeter, $1.8\le\eta\le4.9$, is shown in
fig.\ref{A-eta-forward}, where the dashed, continuous and dotted
lines have the same meaning as before.

Summarizing the large size of the cross section of  $b{\bar
b}b{\bar b}$ production in hadron-nucleus collisions at the LHC
(the values are of the order of one hundreds of $\mu$b,) suggests
that the production of multiple pairs of $b$-quarks is fairly
typical at high energies, hence representing a convenient channel
to study multiple parton interactions. A rather direct feature,
which is a simplest prediction and then a test of the interaction
mechanism described above, is the ``anomalous'' dependence on $A$.
The effects induced by the presence of the nucleonic degrees of
freedom in the nuclear structure are in fact not limited, in this
case, to the shadowing corrections to the nuclear structure
functions usually considered, which cause a limited {\it decrease}
(not larger than $20\%$, in the kinematical regime considered
here) of the cross section for a hard interaction in
hadron-nucleus collisions. When considering double parton
scatterings, all nuclear effects can be exhausted in the shadowing
corrections to the nuclear structure functions only in the
$\sigma_A^D|_1$ term. The dominant effect of the nuclear structure
is however due to the presence of the $\sigma_A^D|_2$ term in the
cross section, which scales with a different power of $A$ as
compared to single scattering term, giving rise to a sizably
larger correction, with opposite sign as compared to the shadowing
correction, namely to an {\it increase} of the cross section,
which may become larger than $100\%$ for a heavy nucleus.

\vskip.25in {\bf Acknowledgment} \vskip.15in This work was
partially supported by the Italian Ministry of University and of
Scientific and Technological Researches (MIUR) by the Grant
COFIN2001.

\newpage
\begin{thebibliography}{99}
\bibitem{DelFabbro:2002pw}
A.~Del Fabbro and D.~Treleani,
Phys. Rev. {\bf D66} (2002) 074012 [arXiv:hep-ph/0207311].
\bibitem{Strikman:2001gz}
M.~Strikman and D.~Treleani, Phys.\ Rev.\ Lett.\  {\bf 88}, 031801
(2002) [arXiv:hep-ph/0111468].
\bibitem{Ryskin:2000bz}
M.~G.~Ryskin, A.~G.~Shuvaev and Y.~M.~Shabelski,
Phys.\ Atom.\ Nucl.\  {\bf 64} (2001) 1995 [Yad.\ Fiz.\  {\bf 64}
(2001) 2080] [arXiv:hep-ph/0007238].

\bibitem{Paver:1982yp}
N.~Paver and D.~Treleani,
Nuovo Cim.\ A {\bf 70}, 215 (1982).
\bibitem{Braun:2000ua}
M.~Braun and D.~Treleani,
Eur.\ Phys.\ J.\ C {\bf 18}, 511 (2001) [arXiv:hep-ph/0005078].

\bibitem{Calucci:1999yz}
G.~Calucci and D.~Treleani,
Phys.\ Rev.\ {\bf D60} (1999) 054023.

\bibitem{DelFabbro:2000ds}
A.~Del Fabbro and D.~Treleani, Phys.\ Rev.\ D {\bf 63}, 057901
(2001) [arXiv:hep-ph/0005273].

\bibitem{Amaudruz:1995tq}
P.~Amaudruz {\it et al.}  [New Muon Collaboration],
Nucl.\ Phys.\ B {\bf 441}, 3 (1995) [arXiv:hep-ph/9503291].
\bibitem{Abe:1997bp}
F.~Abe {\it et al.}  [CDF Collaboration], Phys.\ Rev.\ Lett.\
{\bf 79}, 584 (1997).
\bibitem{Abe:1997xk}
F.~Abe {\it et al.}  [CDF Collaboration],
Phys.\ Rev.\ D {\bf 56}, 3811 (1997).
\bibitem{mrs99}
A.D.Martin, R.G.Roberts, W.J.Stirling and R.S.Thorne, Eur.\ Phys.\
J. {\bf C14} (2000) 133.
\bibitem{madgraph}
T.Stelzer and W.F.Long, Comp.Phys.Comm. {\bf 81}, 357 (1994);
\bibitem{helas}
E.Murayama, I.Watanabe and K.Hagiwara, HELAS: HELicity Amplitude
Subroutines for Feynman Diagram Evaluations, KEK report 91-11,
January 1992;
\bibitem{vegas}
G.~P.~Lepage,
J.\ Comput.\ Phys.\ {\bf 27}, 192 (1978).




\end{thebibliography}
\end{document}